# Exploring fluorine chemical evolution in the Galactic disk: the open cluster perspective⋆

S. Bijavara Seshashayana,[1] H. Jönsson,[1] V. D'Orazi,[2,3] N. Sanna,[4] G. Andreuzzi,[5,6] G. Nandakumar,[7] A. Bragaglia,[8] D. Romano,[8] E. Spitoni[9]

[1] Materials Science and Applied Mathematics, Malmö University, SE-205 06 Malmö, Sweden  
e-mail: `shilpa.bijavara-seshashayana@mau.se`
[2] Department of Physics, University of Rome Tor Vergata, via della Ricerca Scientifica 1, 00133 Rome, Italy
[3] INAF-Osservatorio Astronomico di Padova, vicolo dell' Osservatorio 5, 35122 Padova, Italy
[4] INAF-Osservatorio Astrofisico di Arcetri, Largo E. Fermi 5, 50125, Firenze, Italy
[5] INAF-Osservatorio Astronomico di Roma, via Frascati 33, 00178, Monte Porzio Catone, Italy
[6] Fundacíon *Galileo* Galilei - INAF, Rambla José Ana Fernández Pérez 7, 38712, Brena Baja, Tenerife, Spain
[7] Lund Observatory, Division of Astrophysics, Department of Physics, Lund University, 22100 Lund, Sweden
[8] INAF-Osservatorio di Astrofisica e Scienza dello Spazio di Bologna, via Piero Gobetti 93/3, 40129 Bologna, Italy
[9] INAF-Osservatorio Astronomico di Trieste, via G.B.Tiepolo 11, 34131 Trieste, Italy

**ABSTRACT**

*Context.* Open clusters are ideal tools for tracing the abundances of different elements because their stars are expected to have the same age, distance, and metallicity. Therefore, they serve as very powerful tracers for investigating the cosmic origins of elements. This paper expands on a recent study by us, where the element Fluorine was studied in seven previously open clusters, adding six open clusters as well as eight field stars.

*Aims.* The primary objective is to determine the abundance of fluorine (F) to gain insight into its production and evolution. The magnesium (Mg) abundances were derived to categorize the field stars into high and low alpha disk populations. Additionally, cerium (Ce) abundances are determined to better understand the interplay between F and s-process elements. Our goal is to analyze the trend of F abundances across the Galactic disk based on metallicity and age. By comparing observational data with Galactic chemical evolution models, the origin of F can be better understood.

*Methods.* The spectra were obtained from the high-resolution near-infra-red GIANO-B instrument at the Telescopio Nazionale *Galileo* (TNG). For the derivation of the stellar parameters and abundances, the Python version of Spectroscopy Made Easy (PySME) was used. OH, CN, and CO molecular lines and band heads along with Fe I lines were used to determine the stellar parameters in the H-band region. Two HF lines in the K-band ($\lambda\lambda$ 2.28, 2.33 $\mu$m), three K band Mg I lines ($\lambda\lambda$ 2.10, 2.11, 2.15 $\mu$m), and two Ce II lines in the H-band ($\lambda\lambda$ 1.66, and 1.71 $\mu$m) were used to derive the abundances of F, Mg and Ce, respectively.

*Results.* F, Mg, and Ce abundances were derived for 14 stars from 6 OCs, as well as 8 field stars. The F and Ce abundances were investigated as a function of metallicity, age, and Galactocentric distances. We also compared our findings with different Galactic chemical evolution models.

*Conclusions.* Our results indicate that asymptotic giant branch stars and massive stars, including a subset of fast rotators (whose rotation speed likely increases as metallicity decreases), are necessary to explain the cosmic origin of F. This finding is consistent with and reinforces our previous study, and the larger sample size enhances the strength of the conclusion.

**Key words.** stars: abundances − open clusters and associations: general − Galaxy: evolution − Galaxy: disk

## 1. Introduction

The study and understanding of the cosmic origins of various elements is of crucial importance for solving the mysteries of the formation and evolution of stars as well as understanding the Galactic chemical evolution of these elements. Fluorine (F), an element with an uncertain, yet interesting and intricate origin, will be discussed in this text. Although F has been suggested to be produced by several sources, its exact origin remains a mystery. Therefore, studying the abundance of F in the solar vicinity and beyond is of great importance. Fluorine possibly originates from one or several of five sources: helium-rich intershells in thermally pulsating asymptotic giant branch (TP-AGB) stars (Busso et al. 1999), the helium-burning cores of massive Wolf-Rayet (W-R) stars with strong stellar winds (Meynet & Arnould 2000), primary F from rotating massive stars (Prantzos et al. 2018), through the release of neu-trino flux in Type II supernovae (Woosley & Haxton 1988), and/or novae (José & Hernanz 1998). The main contributor among these five sources is still uncertain. The study of F has experienced growth in recent years due to the utilization of high-resolution IR spectrometers, despite the observational challenges. The metal-poor region ([Fe/H] < -0.8 dex), which according to models should be dominated by F production via the $\nu$ process and rotating massive stars, is not very well studied due to observational challenges, with the lines of interest becoming very weak.

---

⋆ Based on observations made with the Italian Telescopio Nazionale *Galileo* (TNG) operated on the island of La Palma by the Fundación *Galileo* Galilei of the INAF (Istituto Nazionale di Astrofisica) at the Observatorio del Roque de los Muchachos.

Nevertheless, after conducting several observational studies in different environments and comparing them with theoretical models, the five possible contributors were suggested. Jorissen et al. (1992) studied K and M-giants together with AGB stars to determine the abundance of F. The abundance of F for the C-rich AGB stars in this sample was later revised to lower values (Abia et al. 2009; Abia et al. 2010). Still, Jorissen et al. (1992) concluded that AGB stars are the major contributors of Galactic F and that it is produced in the He-burning regions. Carbon-enhanced metal-poor stars (CEMP) were later found to have high F (Schuler et al. 2007; Lucatello et al. 2011). These studies concluded that the high abundances of F in CEMP stars are due to contamination from the previous AGB companions. F has also been detected in helium stars (EHe) (Pandey 2006) and R Corona Borealis variable stars (RCrB) (Pandey et al. 2008), where F was found to be overabundant compared to solar values. Since these objects are thought to be in the AGB phase, this could be a potential source of F. Recio-Blanco et al. (2012) studied F in dwarf stars in the solar neighborhood, D'Orazi et al. (2013) in the globular cluster M22, Werner et al. (2005) in a hot post-AGB star, Zhang & Liu (2005) in planetary nebulae (PNe), and it was concluded that AGB stars are the major contributors. While there is evidence that AGB stars produce F, other sources among the five mentioned may also be involved, leading to some competing notions. Cunha et al. (2003) analyzed F abundances in the Large Magellanic Cloud (LMC) indicating that AGB stars did *not* play a predominant part. An investigation in the Milky Way bulge concluded that F is formed from a mixture of AGB and WR stars (Cunha et al. 2008). Theoretical works by Spitoni et al. 2018; Grisoni et al. 2020 suggest that AGB stars are one possible source, but not necessarily the dominant one. Many EHe stars exhibit overabundances of F in a recent study by Bhowmick et al. (2020). When correlated with C, O, and Ne, it was suggested that they were produced by He and CO white dwarf mergers, as modeled by Menon et al. (2013, 2019); Lauer et al. (2019). Although various studies with different objects conclude that AGB stars likely are the essential driving progenitor for the Galactic chemical evolution of F, W-R stars and $\nu$ processes remain viable sources. The plausible astrophysical sites have been included as potential sources in various Galactic Chemical Evolution (GCE) models (Timmes et al. 1995; Kobayashi et al. 2011a; Kobayashi et al. 2011b; Prantzos et al. 2018; Spitoni et al. 2018; Grisoni et al. 2020). Furthermore, many observational studies have suggested multiple cosmic sources for F, e.g.; Jönsson et al. (2014a), Jönsson et al. (2014b), Pilachowski & Pace (2015), Jönsson et al. (2017), Guerço et al. (2019), Ryde et al. (2020).

The cosmic origin of F is challenging to discern and track due to the absence of robust atomic F lines in the optical or infrared portions of the electromagnetic spectrum. The highly ionized and excited atomic lines of F V and F VI, seen between $1050-1150$Å, are only observed in hot stars (85000-150000 K) in the far UV (FUV) (Werner et al. 2005). F I at $6800-7800$Å in optical spectra of stars have only been used to study EHe and RCBr stars, which have high temperatures ($> 6500$ K) (Pandey 2006, Pandey et al. 2008). For cool stars, the only lines available are IR lines from the HF molecule. Its dissociation energy ($D_0$) is 5.87 eV (Sauval & Tatum 1984), which is rather low, making the lines weak for $T_{eff} > 4500$K, when the molecule becomes dissociated. Combined with the fact that the abundance of F in stars generally is already low, this makes it difficult to trace the Galactic F trend.

All the observational studies and models mentioned previously concentrated on field stars. However, open clusters (OCs) offer valuable insight into abundances, stellar chemistry, and the evolution of the Galaxy, as they contain stars with comparable ages, metallicities, and Galactocentric radius ($R_{gc}$). For this reason, they are known as "ideal laboratories" for testing stellar and Galactic formation and evolution models (Friel 1995; Lada & Lada 2003). In general, over the past few decades, OCs have been the subject of several studies (see e.g., Janes 1979; Bragaglia et al. 2008; Friel et al. 2010; Carrera & Pancino 2011; Carbajo-Hijarrubia et al. 2024). More recently, clusters have also been a subject of study in surveys, for instance, Magrini et al. (2017), Romano et al. (2021), Magrini et al. (2023) for *Gaia*−European Southern Observatory survey (*Gaia*−ESO), Spina et al. (2022) for GALactic Archaeology with HERMES (GALAH), and Myers et al. (2022) for the Apache Point Observatory Galactic Evolution Experiment (APOGEE). However, no current or planned survey will be able to determine the F abundances, since they do not cover the K-band. Smaller-scale works like this paper, mapping the F abundances together with the metallicity ([Fe/H]), age, and $R_{gc}$ will therefore be an important source of information for understanding the evolution of Galactic F. The correlation between F abundances and age and/or $R_{gc}$ in OCs has not yet been extensively studied. Nault & Pilachowski (2013) used the PHOENIX spectrometer to determine the abundance of F in the Hyades, NGC 752, and M67 OCs and concluded that multiple sources may contribute to the F abundance in the disk of our Galaxy, and it is unlikely that AGB is the dominant source. A study by Maiorca et al. (2014) in the OC NGC 6404, using CRIRES at VLT emphasized the significance of the contribution of low-mass AGB stars. Holanda et al. (2024) determined the abundances of 20 elements, including F, in the OC NGC 2345. Their stars in the cluster showed a consistent ratio of F to Fe, indicating a sub-solar abundance of F. Because of all the challenges and constraints mentioned above, more theoretical and observational evidence is needed to search for the cosmic budget of F, which makes it both interesting and an active field of research.

In our recent study (Seshashayana et al. 2024) we presented F and Ce abundances in seven OCs. The abundances were presented in relation to metallicities, ages, $R_{gc}$, and compared with GCE models. Our previous result showed a solar pattern in [F/Fe] without any secondary behavior for stars with solar or supersolar metallicity, with a notable exception for NGC 6791. A comparison of F abundances with GCE models led to the conclusion that both AGB stars and massive stars, including a fraction of fast rotators that increase with decreasing metallicity, are necessary to explain the cosmic origin of F. The present paper builds on and extends this previous work by determining F and Ce abundances for six additional OCs: NGC 6939, NGC 7762, NGC 7142, Collinder 110, Berkeley 32, and NGC 2420, thus extending the already largest homogeneous database for the determination of F in OCs. The objective of selecting these OCs was to increase the number of intermediate-age and old clusters and to improve the coverage in $R_{gc}$. This will facilitate a more comprehensive study of the history of F enrichment in both space and time. In these OCs, we can find giants that are sufficiently cool to permit the measurement of F. In addition, we observed six field stars, that were mainly targeted because they have well-determined stellar parameters and hence serve as a test case for the method used to derive stellar parameters. Furthermore, the stars were sufficiently cool to allow the detection of F.

## 2. Observations

The Telescopio Nazionale *Galileo* (TNG), with a diameter of 3.6m, was used to observe our sample of cluster and field stars (program: A47TAC_6) in March and August 2023, using GIARPS. The instrument configuration allows us to simultaneously observe using the HARPS-N spectrograph (R = 115000, $\lambda\lambda = 3800 - 6900$Å, Cosentino et al. 2014) and GIANO-B near-infrared (NIR) spectrograph (R = 50000, $\lambda\lambda = 0.97 - 2.5\mu$m; Oliva et al. 2012a,b; Origlia et al. 2014). However, since we focus on the NIR part of the spectra, the observing times were calculated using the GIANO-B Exposure Time Calculator, to ensure an S/N around 150 in the K-band when observing at an airmass of 1.5 and a typical seeing condition of 1.2 arcsec. During observations, GIANO-B spectra were collected by nodding the star along the slit, with the target alternatively positioned at 1/4 (position A) and 3/4 (position B) of the slit length for the same amount of time, with single AB nodding exposures of 300 seconds. Fast-rotating hot stars were observed for telluric correction. These stars were observed twice per night at different airmasses. The spectra were processed with the GOFIO reduction software (Rainer et al. 2018), and telluric contamination was eliminated using the standard IRAF technique described for example in Ryde et al. (2019). The basic information regarding all objects can be found in Table 1.

## 3. Analysis

The observed spectra were analyzed using `PySME` (Piskunov & Valenti 2017, Valenti & Piskunov 1996, Wehrhahn et al. 2023), following the methodology employed in our previous investigation (Seshashayana et al. 2024). We used one-dimensional spherical MARCS models (Gustafsson et al. 2008) and linelists retrieved from VALD3 (Kupka et al. 1999, Kupka et al. 2000, Piskunov et al. 1995). `PySME` has support for the use of non-LTE departure coefficient grids, which greatly improves the accuracy in the synthesis of many lines (Amarsi et al. 2016, 2018). For the present work, non-LTE synthesis has been used for Mg- and Fe-lines.

The HF lines used here are R9 at $\lambda_{air} = 23358.329$Å and R16 at $\lambda_{air} = 27778.249$Å. Jönsson et al. (2014a) provides a comprehensive review of the vibrational-rotational K-band HF lines. The Ce II lines are at 16595.180Å and 17058.880Å (Corliss 1973). The $\log gf$ was determined astrophysically for Ce II lines, where a reference abundance is set for an element, and the synthetic spectra are fitted by varying the $\log gf$ value for an observed spectrum of the Sun or another reference star, in this case Arcturus (Cunha et al. 2017). These values were subsequently adjusted slightly by Montelius et al. (2022) to be consistent with abundances derived for the same stars in the solar neighborhood using optical spectra. These values by Montelius et al. (2022) were adapted in our study. The lines used to determine Mg abundances are from the K-band multiplet lines at 21059.76Å, 21060.89Å, and 21458.87Å (this line has a $\log gf$ value with an accuracy of grade B+ in the NIST database). The Mg lines available in our spectra at 15740.70Å and 15748.89Å are excluded from consideration due to saturation and a pronounced correlation with $v_{mic}$ (Nandakumar et al. 2023a).

### 3.1. Stellar Parameters

The stellar parameters such as $T_{eff}$, $\log g$, [Fe/H], and microturbulence ($v_{mic}$) are crucial for the analysis of observed spectra. However, analyzing cool stars ($T_{eff} \lesssim 4500$ K) is difficult because many lines become saturated. The method that we employed for our objects is the one of Nandakumar et al. (2023a), which has been developed especially for the cooler stars and was also used in our previous analysis (Seshashayana et al. 2024). In this iterative method, all stellar parameters except for $\log g$ are determined from the H-band spectra using `PySME`, while $\log g$ is found from isochrones. The fitting procedure is done using a selection of $T_{eff}$-sensitive OH lines and metallicity-sensitive Fe I-lines. Since abundances of C, N, and O are coupled via molecular equilibria, we set as free parameters also C and N, using lines from CN and CO. These features serve to constrain the abundances of C and N, as well as aid in fitting the CO and CN blends in the observed spectra. However, the OH-lines are sensitive not only to $T_{eff}$ but also to O abundance. Therefore, the method needs to have a fixed O abundance to constrain the $T_{eff}$ from the OH lines. To do this, Nandakumar et al. (2023a) used functional forms of trends based on Amarsi et al. (2019)'s 3D NLTE O abundances to calculate [O/Fe] for stars in both the high and low alpha disks for all metallicities. This would make [O/Fe] 0.0 for a solar-metallicity low-alpha star and 0.15 for a high-alpha star. After the fit, $\log g$ is set from the Yonsei-Yale (YY) isochrones (Demarque et al. 2004). `PySME` is then once again used to determine the best fit for the selected spectral lines. The values of $T_{eff}$, [Fe/H], $v_{mic}$, C, and N abundances from the previous cycle will be utilized as new starting values for the subsequent iteration. This process continues until the deviation from the previous cycle is close to zero. As our sample includes field stars, it is necessary to determine whether these stars belong to the high or low alpha disk to set the O to a reasonable value. To accomplish this, we determined the Mg abundances. If the stars belong to the high-alpha disk, we recalculated the stellar parameters using the O abundances appropriate for a high-alpha disk star, see Figure 1. Nandakumar et al. (2023a) estimated an expected error of ±100K for $T_{eff}$, ±0.2dex for $\log g$, ±0.1dex for [Fe/H], ±0.1km s$^{-1}$ for $v_{mic}$, ±0.1dex for [C/Fe], and ±0.1dex for [N/Fe] from this method.

## 3.2. Abundances of F, Mg, and Ce

The abundances of F, Mg, and Ce were derived using the line list information provided in Table 3. From the [Mg/Fe] vs. [Fe/H] trend in Fig. 1, two KIC stars (KIC 5182451 and KIC 12252278) were identified as belonging to the high alpha disk population. For these two stars, the O abundance was set following the high alpha disk trend when determining the stellar parameters. The remaining field stars and all stars belonging to the OCs instead were set following the low alpha disk trend. The [Mg/Fe] values obtained for all the stars in the field are in agreement with the corresponding trend obtained for the stars in the solar neighborhood by Nandakumar et al. (2023a). The values of [Mg/Fe], as well as the other determined abundances, are given in Tables 4-6.

To estimate the uncertainties of the derived abundances in the individual stars, we, just as in our previous study, also redetermined the abundances using 100 different values of $T_{eff}$, log g, [Fe/H], and $v_{mic}$ using a random normal distribution around the parameters we obtained. The evaluation was based on the methodological uncertainty estimates by Nandakumar et al. (2023a). Hence, each spectrum was analyzed 100 times using 100 different sets of stellar parameters, yielding 100 different F, Mg, and Ce abundance values. The abundance uncertainties for the individual stars in Tables 4-6 are given as the median absolute deviation of the individual abundance uncertainties for each star.

Further, just as in our previous analysis, two $v_{mac}$ determinations were performed. A smaller $v_{mac}$ has been chosen from the H-band lines and is used for the determination of Ce abundances from the H-band, and a slightly larger value was set for the K-band HF lines. The reason for this is to account for the slight variation in the instrument resolution with wavelength.

## 4. Results and discussion

The main results are presented in Tables 4 - 7 and illustrated in Figures 2 - 4. The relationships between [F/Fe], [Ce/Fe], [F/H], and [Ce/H] with age and $R_{gc}$ were investigated only for our OCs. For field stars, only the relationship between [F/Fe], [Ce/Fe], [F/H], and [Ce/H] with [Fe/H] was explored. The results are compared with Galactic chemical evolution (GCE) models that implement different stellar sources of F in Figure 4.

### 4.1. Comparison of our stars' parameters with literature values.

Six stars were taken from the Kepler database, and have asteroseismic log g-values in the APOKASC-3 catalog (private communication, M. Pinsonneault et al. in prep), that can be used to test the performance of our iterative method to derive log g using spectra and isochrones. The agreement is generally very good, see Table 5.

The two HIP stars have $T_{eff}$ determined from angular diameter measurements (Baines et al. 2021), which can be used to assess our ability to correctly derive $T_{eff}$. The resulting values are well within the error ranges (see Table 6).

The metallicities of all clusters determined by us are compared with a few literature values in Table 2. The Table presents our [Fe/H] values, the corresponding literature values, and the respective literature references for all six clusters. The resultant metallicities are well within the range of error in comparison to previous literature values.

### 4.2. F and Ce abundances

In Figure 2, the metallicity is plotted with [F/Fe] and [F/H] for our OCs and field stars. The results are presented in combination with the field stars from Nandakumar et al. (2023b) in the upper panels. Our results are also plotted with the literature OCs for [Fe/H], age, and $R_{gc}$ (see upper, middle, and lower panels). Additionally, a comparison is presented between the current study and our previous results (seven OCs that we had previously studied; see Seshashayana et al. 2024). The data points align with the literature on OCs and field stars. At super-solar metallicities, F exhibits a secondary behavior in Ryde et al. (2020) and Guerço et al. (2022), which is not present in Nandakumar et al. (2023b). Our results also support those of Nandakumar et al. (2023b), as the stronger and possibly saturated/blended HF R9 line, which shows temperature-dependent trends, is found to have high uncertainty, especially in cooler and metal-rich giant stars (for a detailed explanation, see Nandakumar et al. 2023b). The middle and lower panels are compared with literature OCs that have been previously analyzed for F. Since the age and $R_{gc}$ for field stars are less reliable than for OCs, these panels only contain our clusters and not field stars. Hyades, NGC 752, M67 from Nault & Pilachowski (2013), NGC 2345 from (Holanda et al. 2023), and finally M67 (again) and NGC 6404 from Maiorca et al. (2014) comprises all the OCs that have been employed thus far to investigate F, disregarding our previous paper. Our combined data set from the previous and this paper, together with NGC 2345, show a uniform flat trend between about 200 Myr and 6 Gyr, except for NGC 6791, which shows a high [F/Fe] ratio (see Seshashayana et al. 2024 for more details). For the clusters analyzed by us, there is a downward trend when [F/H] is plotted against age and $R_{gc}$, which is in agreement with standard chemical evolution and the presence of the Galactic metallicity gradient. However, the literature OCs NGC 752, Hyades, and NGC 2345 deviate significantly from the data analyzed by us, rendering it inadvisable to draw any inferences when plotted with our data. Lastly, when [F/Fe] is plotted against age, a flat trend is seen. The OCs from the current analysis follow the trend of our previous OCs. The slope for [F/Fe] versus $R_{gc}$ without NGC 6791 is $0.00 \pm 0.02$ dex/kpc. The value from the previous analysis in Seshashayana et al. (2024) is $0.01 \pm 0.02$ dex/kpc. In our current study, when we calculate the slope for [F/H] versus $R_{gc}$ for all of our clusters except NGC 6791, we find a value of $-0.09 \pm 0.02$ dex/kpc. The slope from our previous analysis for [F/H] versus $R_{gc}$ (see Seshashayana et al. 2024) is $-0.09 \pm 0.03$ dex/kpc. The slope value provided by the current sample of stars compared to our previous results indicates that the data are consistent across a wider range of stars, providing a more comprehensive and robust understanding of the behavior of F.

The same was done for Ce, where [Ce/Fe] and [Ce/H] were compared with [Fe/H], age, and $R_{gc}$. Figure 3 shows the behavior of Ce for our clusters. We compared with literature OCs from Myers et al. (2022) and some field stars from Ryde et al. (2020). The OCs from Myers et al. (2022)

| Stellar cluster | Star | *Gaia* DR3 ID | RA (deg) | Dec (deg) | G (mag) | S/N H-band | S/N K-band |
|---|---|---|---|---|---|---|---|
| NGC 6939 | NGC 6939_1 | 2194725337620737792 | 307.690027985254 | 60.60090033682878 | 10.34 | 249 | 216 |
| age=1.70Gyr | NGC 6939_2 | 2194824018782231296 | 307.751916928197 | 60.81436876801729 | 11.12 | 224 | 199 |
| $R_{gc} = 8.70$ kpc | NGC 6939_3 | 2194726471490897024 | 307.518549581646 | 60.55418772365226 | 11.19 | 148 | 132 |
|  | NGC 6939_4 | 2194713139912733056 | 307.918942123322 | 60.61892205660228 | 11.32 | 166 | 145 |
| NGC 2420 | NGC 2420_1 | 865399939794984320 | 114.56276636 | 21.5830497562 | 11.14 | 198 | 172 |
| age=1.74Gyr |  |  |  |  |  |  |  |
| $R_{gc} = 10.68$ kpc |  |  |  |  |  |  |  |
| Collinder 110 | Cr 110_2 | 3126738188653591168 | 99.805185271 | 2.0493381003 | 11.05 | 230 | 194 |
| age=1.82Gyr | Cr 110_3 | 3126749767885447552 | 99.738812533 | 2.05903237577 | 10.98 | 254 | 230 |
| $R_{gc} = 10.29$ kpc | Cr 110_4 | 3126754234651093888 | 99.686343516 | 2.14405637405 | 10.19 | 390 | 326 |
| NGC 7762 | NGC 7762_1 | 2210930249228141056 | 357.800686711477 | 67.94158562607588 | 10.41 | 262 | 230 |
| age=2.04Gyr |  |  |  |  |  |  |  |
| $R_{gc} = 8.78$ kpc |  |  |  |  |  |  |  |
| NGC 7142 | NGC 7142_1 | 2217941937956979840 | 326.260545852642 | 65.76114212449832 | 11.26 | 186 | 162 |
| age=3.09Gyr | NGC 7142_2 | 2217944274419272960 | 326.705463196390 | 65.76928923282846 | 11.71 | 171 | 151 |
| $R_{gc} = 9.25$ kpc | NGC 7142_3 | 2217943587224390912 | 326.291673041744 | 65.85459299930302 | 12.01 | 132 | 112 |
| Berkeley 32 | Be 32_1 | 3129900070557039488 | 104.51412308 | 6.44718036949 | 11.66 | 254 | 230 |
| age=4.90Gyr | Be 32_2 | 3129898833606459904 | 104.51319996 | 6.40616048404 | 11.80 | 148 | 115 |
| $R_{gc} = 11.14$ kpc |  |  |  |  |  |  |  |
|  | HIP 3786 | 2556759229189045632 | 12.170968944689 | 7.5848586637986 | 3.84 | 438 | 395 |
|  | HIP 4147 | 2535891460567764224 | 13.252085422292 | -1.1443296982576 | 4.12 | 493 | 383 |
|  | KIC 5182451 | 2101176414429414528 | 290.400321483165 | 40.3197644483823 | 10.54 | 87 | 78 |
|  | KIC 8414116 | 2106673762107078656 | 284.096587350876 | 44.4797275017186 | 10.09 | 219 | 190 |
|  | KIC 8427166 | 2126971850569328128 | 290.462597612610 | 44.4461778834982 | 10.69 | 185 | 169 |
|  | KIC 12252278 | 2133370527203207040 | 287.902989291060 | 50.9613014077372 | 10.69 | 142 | 123 |
|  | KIC 11076239 | 2131204145699718528 | 287.619303757366 | 48.6328776694531 | 10.25 | 136 | 120 |
|  | KIC 10801138 | 2128855726303447296 | 293.429473531675 | 48.1947937842640 | 10.00 | 202 | 176 |

Table 1: Basic information for our program stars, including the eight field stars, and their spectra. The ages and $R_{gc}$ for the OCs are from Cantat-Gaudin et al. (2020). The header of each file containing the spectrum information was employed to obtain the signal-to-noise ratios (S/N). This was achieved by utilizing the value for order 48 as a reference for the H-band and order 33 for the K-band.

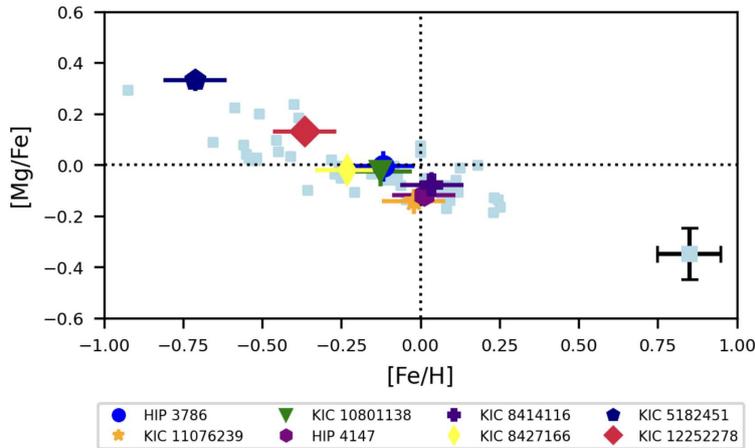

Fig. 1: [Mg/Fe] versus [Fe/H]. The field stars were plotted to categorize high and low alpha disk stars. The light blue squares are taken from Nandakumar et al. (2023a). Representative uncertainties are shown in the bottom-right corner.

| Stellar Cluster | [Fe/H] determined in this article (dex) | Literature [Fe/H] Value (dex) | Literature Reference |
|---|---|---|---|
| NGC 6939 | $-0.01 \pm 0.01$ | $-0.19 \pm 0.1$ | Friel et al. (2002) |
|  |  | $0.00 \pm 0.1$ | Jacobson et al. (2007) |
|  |  | $-0.02 \pm 0.15$ | Overbeek et al. (2016) |
| NGC 7142 | $0.04 \pm 0.01$ | $-0.10 \pm 0.1$ | Friel et al. (2002) |
|  |  | $0.08 \pm 0.06$ | Jacobson et al. (2007) |
| NGC 7762 | $-0.03 \pm 0.04$ | $0.04 \pm 0.12$ | Carraro et al. (2016) |
| Berkeley 32 | $-0.22 \pm 0.08$ | $-0.29 \pm 0.1$ | Sestito et al. (2006) |
|  |  | $-0.21 \pm 0.01$ | Friel et al. (2010) |
|  |  | $-0.26 \pm 0.06$ | Overbeek et al. (2016) |
|  |  | $-0.31 \pm 0.2$ | Zhong et al. (2020) |
|  |  | $-0.38 \pm 0.06$ | Spina et al. (2021) |
|  |  | $-0.31 \pm 0.06$ | Randich et al. (2022) |
|  |  | $-0.37 \pm 0.2$ | Fu et al. (2022) |
| Collinder 110 | $-0.10 \pm 0.04$ | $0.04 \pm 0.05$ | Pancino et al. (2010) |
|  |  | $-0.1 \pm 0.04$ | Randich et al. (2022) |
| NGC 2420 | $-0.16 \pm 0.04$ | $-0.20 \pm 0.06$ | Jacobson et al. (2011) |
|  |  | $-0.05 \pm 0.05$ | Pancino et al. (2017) |
|  |  | $-0.12 \pm 0.02$ | Donor et al. (2018) |
|  |  | $-0.12 \pm 0.03$ | Carrera et al. (2019) |
|  |  | $-0.19 \pm 0.03$ | Spina et al. (2021) |
|  |  | $-0.20 \pm 0.04$ | Myers et al. (2022) |
|  |  | $-0.15 \pm 0.02$ | Randich et al. (2022) |

Table 2: Comparison of our metallicity values with those found in the literature. Each value is referenced.

are from the APOGEE survey (DR17, Abdurro'uf et al. 2022), and include 150 OCs. However, we extracted clusters with available data on ages, distances, and [Ce/Fe] values, resulting in a subsample of 104 OCs for head-to-head comparison with our results. Our current analysis shares one cluster, NGC 2420, and from our previous analysis, four clusters (NGC 6791, NGC 6819, Trumpler 5, and NGC 7789) with Myers et al. (2022). The upper panels show that our obtained values are in good general agreement with those reported by both Myers et al. (2022) and Ryde et al. (2020). From our current analysis, the [Ce/Fe] value of NGC 2420 is $0.30 \pm 0.09$ and from Myers et al. (2022) it is $0.16 \pm 0.14$, which makes our value higher, although within the observational uncertainty. For [Ce/Fe] relative to age, there is a flat trend, while for $R_{gc}$ there is an increasing trend. For [Ce/H] relative to age and $R_{gc}$, there is a decreasing trend. This suggests that younger clusters tend to have higher Ce abundance compared to their older counterparts. This finding is in line with previous studies on s-process elements in OCs, indicating a rising trend in younger clusters (D'Orazi & Randich 2009, Maiorca et al. 2011, Yong et al. 2012, Mishenina et al. 2015, D'Orazi et al. 2022, Magrini et al. 2023). The increasing trend of [Ce/Fe] concerning $R_{gc}$ confirms previous findings on OCs in for example Gaia-ESO, such as those by Viscasillas Vázquez et al. (2022) and Magrini et al. (2023). The calculated slope for all our clusters is given in Fig. 3 and the value is $0.03 \pm 0.01$ dex/kpc (without NGC 6791) which is in good agreement with the findings of Myers et al. (2022) who found $0.02 \pm 0.01$ dex/kpc. Our analysis also shows the value for [Ce/H] versus $R_{gc}$ which is $-0.02 \pm 0.01$ dex/kpc without NGC 6791. The value from our previous results (see Seshashayana et al. (2024)) for [Ce/Fe] versus $R_{gc}$ is $0.03 \pm 0.03$ dex/kpc and [Ce/H] versus $R_{gc}$ is $-0.07 \pm 0.05$ dex/kpc. A comparison of the slope value provided by the current sample of stars for Ce analysis with the results of previous studies indicates that the data is consistent with the results of the previous study.

In Figure 4 we compare our results with chemical evolution models that investigate F to put constraints on its cosmic origin. The different production channels come from various major sources, such as rotating and non-rotating massive stars, AGB stars, W-R stars, SNe II, and Novae, which are suggested to produce the F that we observe today. Figure 4 shows [F/Fe], [F/H] against [Fe/H] and age. We compare the models with the stars from our previous analysis, some literature OCs, and field stars (similar to Fig. 2).

The galactic chemical evolution models utilized for comparison in the current study are the same that were developed for our previous paper, Seshashayana et al. (2024). There are three sets of yields considered in this model. The massive star yields (including W-R stars) with different rotation rates are assumed to have no rotation ($v_{rot} = 0$ km/s) (Limongi & Chieffi 2018), and fast-rotating stars with rotation rates ($v_{rot} = 150$ and $300$ km/s). The yields of low- and intermediate-mass stars (LIMS), such as AGB stars are taken from either the database the FUll-Network Repository of Updated Isotopic Tables and Yields (FRUITY) database (Cristallo et al. 2009, 2011, 2015) or the yields by the Monash team (Lugaro et al. 2012; Fishlock et al. 2014; Karakas & Lugaro 2016; Karakas et al. 2018). For a detailed description of the yields chosen, please refer to our previous paper, Seshashayana et al. (2024).

Based on the data presented in Figure 4, it is clear that the AGB F yields from MONASH are sufficient to roughly explain the F abundances observed, but massive star yields where F is produced in the hydrostatic burning layers are needed when FRUITY yields are considered. The F yields for rotating massive stars from Limongi & Chieffi (2018) result in excessive production of F. Previous studies (Prantzos et al. 2018, Romano et al. 2019) have shown that the $v_{rot}$ of massive stars must be faster at low metallicities, and vice versa. A new model was introduced in Seshashayana et al. (2024, cyan solid line in Fig. 4). In this model, 80% of low-metallicity massive stars are fast rotators, while the remaining 20% are considered to not rotate. The ratio is the other way around when considering solar and supersolar metallicities. A gradual transition exists between the two ranges. The model where yields from LIMS (MONASH) and massive stars are mixed like this, is the best fit of the observations within the error range.

## 5. Summary and Conclusions

This paper presents a study of stars in six open clusters and eight field stars. It also presents the values of stellar parameters, metallicities, fluorine, magnesium, and cerium abundances. The sample is composed of a total of 22 stars, with 14 stars spread across the six open clusters. The open clusters under consideration are NGC 7142, NGC 6939, NGC 7762, Collinder 110, Berkeley 32, and NGC 2420. The objective of this study was to investigate the cosmic origin of F and Ce by analyzing their trends. The study compared the abundances of F and Ce with metallicity for both open clusters (OCs) and field stars, as well as with literature values. The abundances of F and Ce for OCs were plotted against age and $R_{gc}$ to identify global trends. Our sample shows a flat trend when plotting [F/Fe] with metallicity, age, and $R_{gc}$. On the other hand, a downward trend is observed for [F/H] and [Ce/H] versus age and $R_{gc}$, which is consistent with the expected global Galactic gradient. Our results have been compared with models that assume different F production channels, including low-mass stars with yields from FRUITY (Cristallo et al. 2009, Cristallo et al. 2011, Cristallo et al. 2015) as well as MONASH (Lugaro et al. 2012, Fishlock et al. 2014, Karakas & Lugaro 2016, Karakas et al. 2018), and rotating massive stars Limongi & Chieffi (2018). The FRUITY-LIMS are underproducing F, indicating the need for another source. The abundances are overestimated by rapidly rotating massive stars. However, the yields from MONASH-LIMS, provide a reasonable fit to our data. On the other hand, when considering age, it is evident that an additional source is still necessary. To better explain the observational results, we considered AGB stars (MONASH yields) and a proportion of fast rotators, which varies as a function of metallicity i.e., there ought to be more fast rotators on average with decreasing metallicity (Prantzos et al. 2018, Romano et al. 2019). Although the models provide a satisfactory explanation of the origin of F, a larger sample with a wider variety of ages and galactocentric distances would be beneficial. In particular, it would be advantageous to include younger and older OCs, thereby expanding the current sample. Nevertheless, it is crucial to acknowledge that the majority of open clusters observed are not sufficiently older than those included in our present sample. Moreover, the cool giants necessary for the current analysis are not present in younger open clusters. For this reason, it would be beneficial to investigate cool dwarfs in such clusters. Nevertheless, the current results strengthen our previous claim that AGB stars alone cannot account for the production of F. Other sources, such as massive stars with varying rotation rates, must also be involved.

*Acknowledgements.* S.B.S. acknowledges funding from the Crafoord Foundation. A.B. acknowledges funding from INAF MiniGrant 2022. V.D. acknowledges the financial contribution from PRIN MUR 2022 (code 2022YP5ACE) funded by the European Union – NextGenerationEU.

| Species | $\lambda_{air}$ (Å) | log($gf$) | $E_{low}$ (eV) | Reference |
|---|---|---|---|---|
| HF | 22778.249 | -3.969 | 0.674 | Jönsson et al. (2014a) |
| HF | 23358.329 | -3.962 | 0.227 | Jönsson et al. (2014a) |
| Mg I | 21059.757 | -0.384 | 6.779 | Civiš et al. (2013) |
| Mg I | 21060.896 | -1.587 | 6.779 | Civiš et al. (2013) |
| Mg I | 21060.896 | -0.407 | 6.779 | Civiš et al. (2013) |
| Mg I | 21458.962 | -1.319 | 6.516 | Nandakumar et al. (2023a) |
| Ce II | 16595.180 | -2.114 | 0.122 | Montelius et al. (2022) |
| Ce II | 17058.880 | -1.425 | 0.318 | Montelius et al. (2022) |

Table 3: Lines used to determine magnesium, fluorine, and cerium abundances.

| Star | $T_{eff}$ (K) | log$g$ (dex) | [Fe/H] (dex) | $v_{mic}$ (km s$^{-1}$) | $v_{mac}$ (km s$^{-1}$) | [C/Fe] (dex) | [N/Fe] (dex) | [O/Fe] (dex) | [Mg/Fe] (dex) | [F/Fe] (dex) | [Ce/Fe] (dex) |
|---|---|---|---|---|---|---|---|---|---|---|---|
| NGC 6939_1 | 3980 | 1.53 | 0.00 | 1.67 | 8.15 | -0.17 | 0.23 | -0.04 | -0.12±0.04 | -0.13±0.12 | 0.06±0.08 |
| NGC 6939_2 | 4106 | 1.77 | 0.00 | 1.53 | 8.58 | -0.16 | 0.23 | -0.04 | -0.14±0.05 | -0.17±0.14 | 0.19±0.10 |
| NGC 6939_3 | 4285 | 2.10 | -0.03 | 1.68 | 8.13 | -0.13 | 0.18 | -0.03 | -0.08±0.04 | -0.29±0.13 | 0.12±0.07 |
| NGC 6939_4 | 4297 | 2.12 | -0.02 | 1.56 | 7.13 | -0.14 | 0.26 | -0.03 | -0.14±0.06 | -0.18±0.14 | 0.18±0.08 |
| NGC 2420_1 | 4083 | 1.61 | -0.16 | 1.60 | 7.03 | -0.21 | 0.22 | 0.02 | -0.05±0.04 | -0.16±0.16 | 0.30±0.09 |
| Cr 110_2 | 4026 | 1.56 | -0.08 | 1.51 | 7.78 | -0.17 | 0.19 | -0.01 | -0.08±0.06 | -0.33±0.15 | 0.23±0.09 |
| Cr 110_3 | 4005 | 1.48 | -0.13 | 1.62 | 8.54 | -0.12 | 0.19 | 0.01 | -0.08±0.04 | -0.23± 0.15 | 0.17±0.07 |
| Cr 110_4 | 3701 | 0.95 | -0.08 | 1.90 | 7.87 | -0.16 | 0.28 | -0.01 | -0.07±0.05 | -0.25±0.12 | 0.05±0.07 |
| NGC 7762_1 | 4201 | 1.93 | -0.03 | 1.59 | 7.67 | -0.14 | 0.24 | -0.03 | -0.15±0.05 | -0.15±0.16 | 0.25±0.06 |
| NGC 7142_1 | 3942 | 1.49 | 0.05 | 1.75 | 8.95 | -0.13 | 0.21 | -0.06 | -0.03±0.04 | -0.19±0.12 | -0.10±0.06 |
| NGC 7142_2 | 4072 | 1.72 | 0.04 | 1.66 | 7.62 | -0.16 | 0.22 | -0.05 | -0.11±0.04 | -0.27±0.13 | -0.06±0.09 |
| NGC 7142_3 | 4130 | 1.83 | 0.03 | 1.45 | 8.06 | -0.15 | 0.30 | -0.05 | -0.16±0.03 | -0.24±0.16 | 0.02±0.10 |
| Be 32_1 | 4007 | 1.48 | -0.14 | 1.61 | 6.67 | -0.13 | 0.19 | 0.01 | -0.09± 0.06 | -0.23±0.17 | 0.19±0.09 |
| Be 32_2 | 4132 | 1.60 | -0.31 | 1.50 | 6.52 | -0.15 | 0.23 | 0.07 | -0.08±0.05 | -0.07±0.20 | 0.09±0.08 |

Table 4: Stellar parameters and abundances for the OC stars. Abundance ratios are scaled to the solar values of A(C)$_\odot$ = 8.46, A(N)$_\odot$ = 7.83, A(O)$_\odot$ = 8.69, A(Mg)$_\odot$ = 7.55, A(F)$_\odot$ = 4.40, A(Fe)$_\odot$ = 7.46, and A(Ce)$_\odot$ = 1.58 (Asplund et al. 2021) . Typical uncertainties of our derived stellar parameters are ±100 K in T$_{eff}$, ±0.2 dex in log$g$, ±0.1 dex in [Fe/H], ±0.1 km s$^{-1}$ in v$_{mic}$, ±0.1 dex in [C/Fe], and ±0.1 dex in [N/Fe]. The values given here for v$_{mac}$ are derived from the H-band.

| Star | $T_{eff}$ (K) | $\log g$ (dex) | $\log g_A$ (dex) | [Fe/H] (dex) | $v_{mic}$ (km s$^{-1}$) | $v_{mac}$ (km s$^{-1}$) | [C/Fe] (dex) | [N/Fe] (dex) | [O/Fe] (dex) | [Mg/Fe] (dex) | [F/Fe] (dex) | [Ce/Fe] (dex) |
|---|---|---|---|---|---|---|---|---|---|---|---|---|
| KIC 5182451 | 3776 | 0.73 | 0.64 | -0.67 | 1.96 | 6.73 | 0.05 | 0.06 | 0.43 | 0.33±0.04 | -0.19±0.12 | -0.12±0.07 |
| KIC 8414116 | 3991 | 1.45 | 1.44 | 0.05 | 1.48 | 8.23 | -0.16 | 0.12 | -0.06 | -0.08±0.05 | -0.13±0.11 | -0.14±0.08 |
| KIC 8427166 | 3556 | 0.63 | 0.58 | -0.21 | 1.77 | 8.84 | -0.13 | 0.06 | 0.03 | -0.02±0.05 | 0.11±0.09 | -0.15±0.10 |
| KIC 10801138 | 3997 | 1.46 | 1.33 | -0.19 | 1.57 | 7.96 | -0.16 | 0.23 | 0.00 | -0.03±0.06 | -0.09±0.14 | 0.04±0.08 |
| KIC 11076239 | 4018 | 1.60 | 1.42 | 0.00 | 1.63 | 7.77 | -0.15 | 0.17 | -0.05 | -0.14±0.05 | -0.03±0.16 | -0.08±0.07 |
| KIC 12252278 | 4021 | 1.35 | 1.29 | -0.35 | 1.50 | 8.69 | -0.03 | 0.14 | 0.28 | 0.13±0.05 | -0.30±0.13 | -0.10±0.09 |

Table 5: Stellar parameters and abundances for all the Kepler field stars. Our derived $\log g$ values have been compared with asteroseismic $\log g$ values that are represented as $\log g_A$. The uncertainties of $\log g_A$ are all within 0.01 dex. For solar reference values, uncertainties, and $v_{mac}$ of our values, see caption of Table 4.

| Star | $T_{eff}$ (K) | $T_{teff,A}$ (K) | $\log g$ (dex) | [Fe/H] (dex) | $v_{mic}$ (km s$^{-1}$) | $v_{mac}$ (km s$^{-1}$) | [C/Fe] (dex) | [N/Fe] (dex) | [O/Fe] (dex) | [Mg/Fe] (dex) | [F/Fe] (dex) | [Ce/Fe] (dex) |
|---|---|---|---|---|---|---|---|---|---|---|---|---|
| HIP 3786 | 3927 | 3868±35 | 1.36 | -0.11 | 1.61 | 6.75 | -0.11 | 0.16 | -0.01 | -0.01±0.06 | -0.08±0.12 | -0.17±0.10 |
| HIP 4147 | 3883 | 3724±35 | 1.36 | 0.02 | 1.60 | 8.57 | -0.22 | 0.16 | -0.05 | -0.12±0.04 | -0.19±0.12 | 0.24±0.08 |

Table 6: Stellar parameters and abundances for the field stars with measured angular diameters from Baines et al. (2021) that is represented as $T_{teff,A}$. For solar reference values, uncertainties, and $v_{mac}$ of our values, see caption of Table 4.

| Stellar cluster | Age (Gyr) | $R_{GC}$ (kpc) | [Fe/H] (dex) | [Mg/Fe] (dex) | [Mg/H] (dex) | [F/Fe] (dex) | [F/H] (dex) | [Ce/Fe] (dex) | [Ce/H] (dex) |
|---|---|---|---|---|---|---|---|---|---|
| NGC 6939 | 1.70 | 8.70 | -0.01±0.01 | -0.12±0.02 | -0.13±0.02 | -0.19±0.04 | -0.21±0.04 | 0.14±0.04 | 0.12±0.05 |
| NGC 2420 | 1.74 | 10.68 | -0.16±0.04 | -0.05±0.03 | -0.21±0.06 | -0.16±0.04 | -0.31±0.03 | 0.30±0.04 | 0.14±0.07 |
| Collinder 110 | 1.82 | 10.29 | -0.10±0.04 | -0.08±0.01 | -0.17±0.03 | -0.27±0.02 | -0.40±0.01 | 0.14±0.05 | 0.04±0.08 |
| NGC 7762 | 2.04 | 8.78 | -0.03±0.04 | -0.15±0.03 | -0.18±0.06 | -0.15±0.04 | -0.18±0.03 | 0.16±0.04 | 0.13±0.07 |
| NGC 7142 | 3.09 | 9.25 | 0.04±0.01 | -0.10±0.07 | -0.06±0.07 | -0.23±0.03 | -0.19±0.01 | -0.05±0.04 | -0.01±0.03 |
| Berkeley 32 | 4.90 | 11.14 | -0.22±0.08 | -0.09±0.01 | -0.31±0.12 | -0.15±0.08 | -0.37±0.04 | -0.04±0.05 | -0.26±0.14 |

Table 7: The six clusters' averaged abundance values are presented. The uncertainties mentioned above are computed from the root mean square deviation from the mean of the values obtained for individual stars within the cluster, except for NGC 7762 and NGC 2420, where we have only one star in our sample. For these clusters, the reported uncertainties are instead representative measures that are based on the average uncertainties of the other clusters.

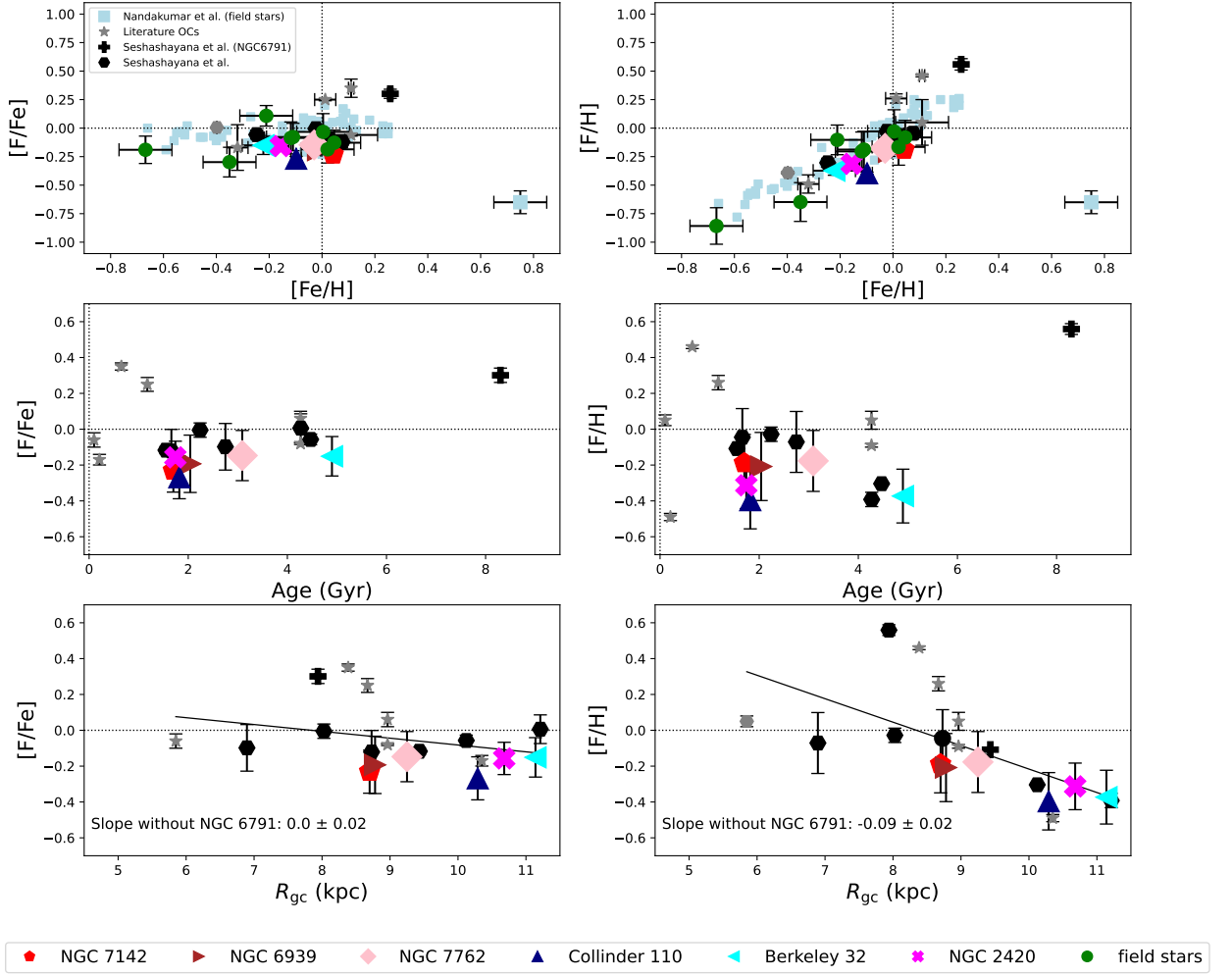

Fig. 2: Left: Relationship between [F/Fe] and [Fe/H], age, and $R_{gc}$. Right: Same but for [F/H]. Field stars from Nandakumar et al. (2023b) are depicted as light blue squares. A representative error bar for the data from Nandakumar et al. (2023b) is shown in the bottom right corner in the first row of panels. The black hexagons are the OCs studied in our previous paper, Seshashayana et al. (2024). We also plot values for five OCs as light gray stars, namely Hyades, NGC 752, and M67 from Nault & Pilachowski (2013), another estimate of M67 and another cluster, NGC 6404, from Maiorca et al. (2014), and NGC 2345 from Holanda et al. (2023). The slopes shown in the figure were obtained by a fit to all strictly similarly analyzed OCs from this and our previous work, excluding the outlier NGC6791 (see Seshashayana et al. 2024, for discussion).

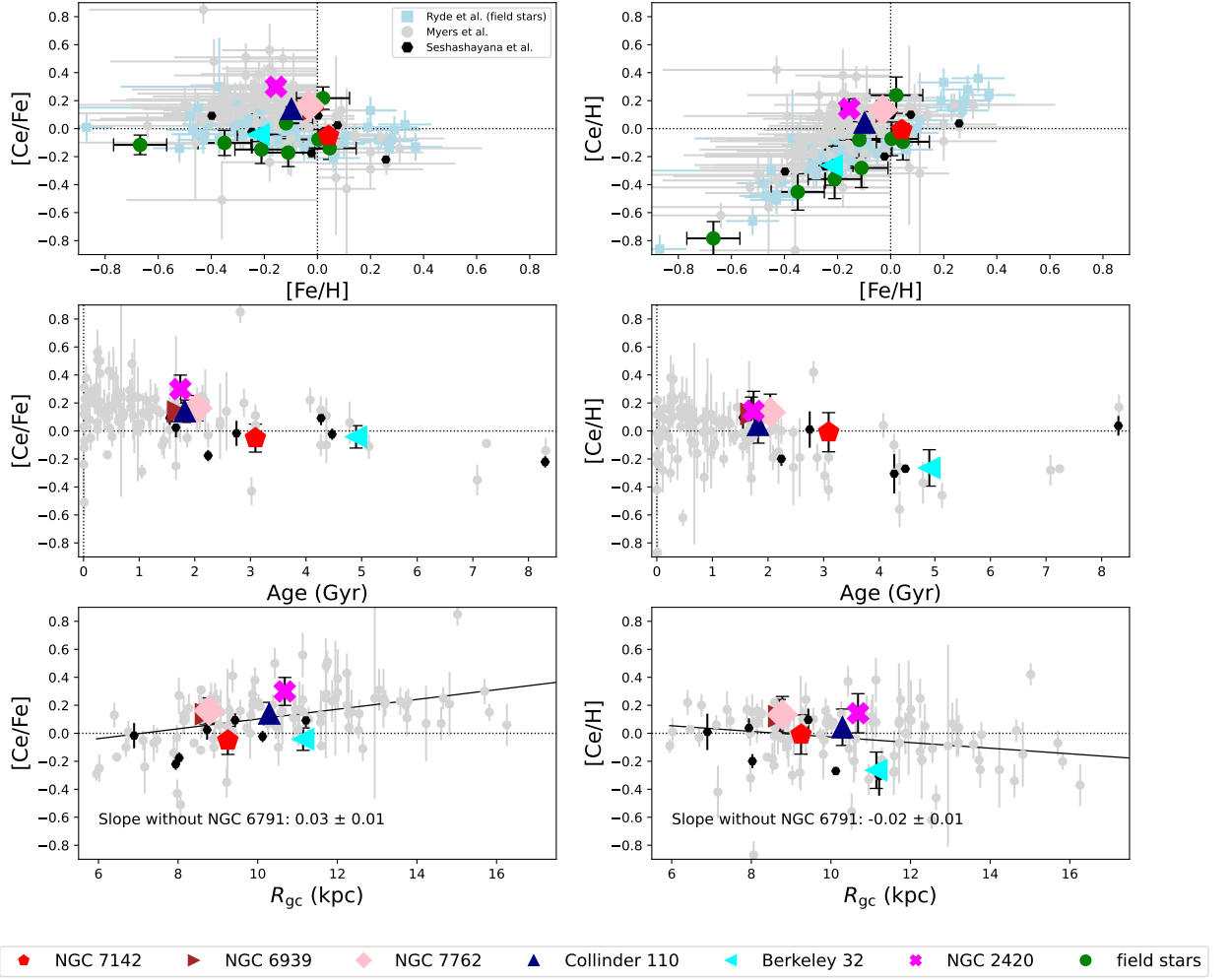

Fig. 3: Relationship between [Ce/Fe] and [Fe/H], age, and $R_{gc}$ (left) and [Ce/H] as a function of [Fe/H], age, and $R_{gc}$ (right). The black hexagons are the OCs studied in our previous paper, Seshashayana et al. (2024). Data for field stars taken from Ryde et al. (2020) are shown as light blue dots. The clusters taken from Myers et al. (2022) are shown as light gray squares. The slopes shown in the figure were obtained by a fit to all strictly similarly analyzed OCs from this and our previous work, excluding the outlier NGC6791 (see Seshashayana et al. 2024, for discussion).

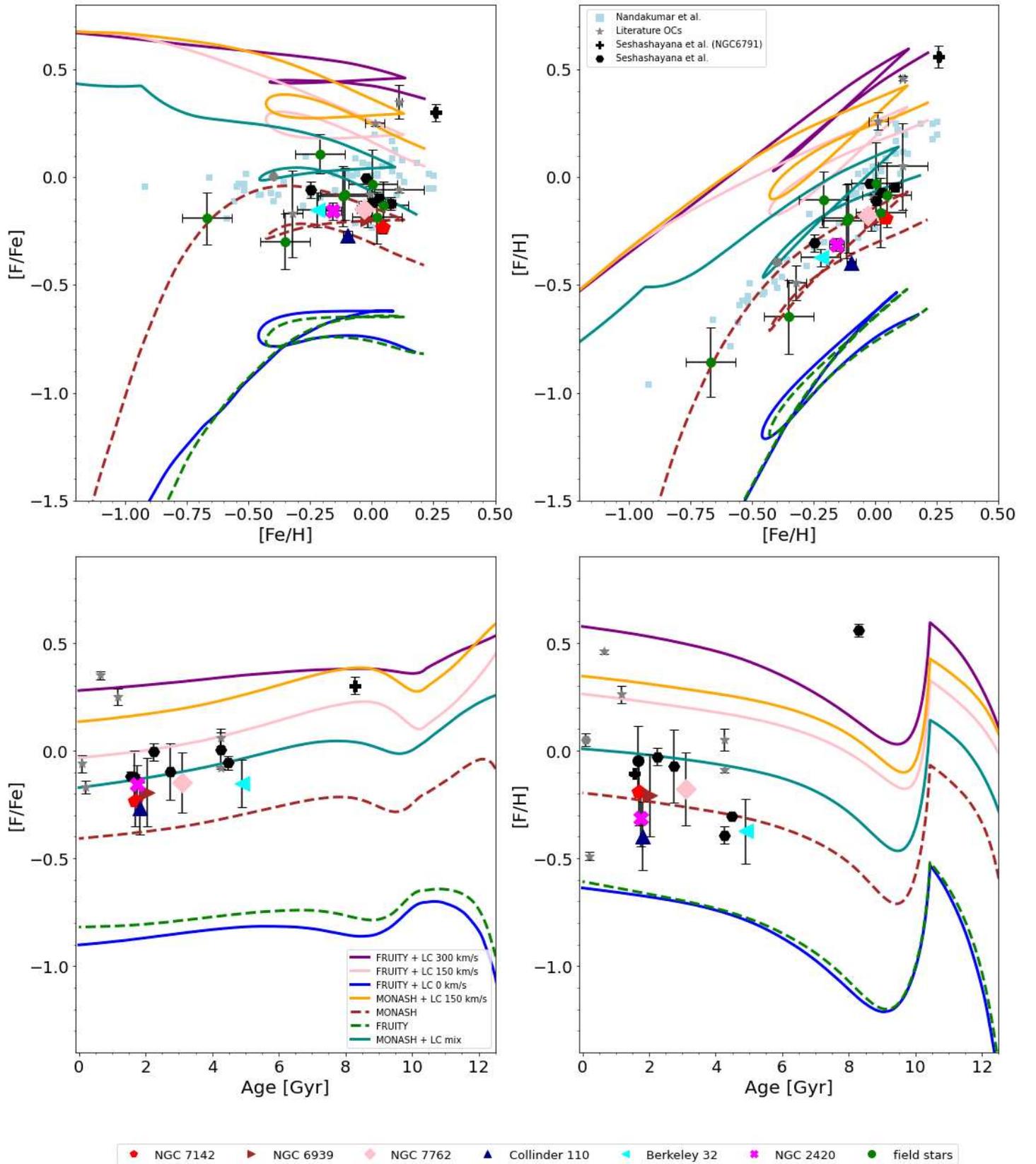

Fig. 4: [F/Fe] and [F/H] against [Fe/H] (top) and age (bottom). The data points are measurements for OCs as well as field stars, and the lines are predicted from theoretical models. The OCs from our previous sample are shown as black hexagons. The figure shows the models that allow F to form from both LIMS and massive stars (solid lines) and the models that only allow F to form from AGB stars (dashed lines).